\documentclass[twoside]{article}
\usepackage[accepted]{aistats2e}
\usepackage{amsmath,amssymb,graphicx,natbib}

\def\bbeta{\boldsymbol{\beta}}

\def\bphi{\boldsymbol{\phi}}

\def\bc{\boldsymbol{c}}

\def\bn{\boldsymbol{n}}

\def\bu{\boldsymbol{u}}   
\def\bw{\boldsymbol{w}}

\def\bH{\boldsymbol{H}}

\def\bX{\boldsymbol{X}}

\def\g{\,|\,}

\begin{document}

\runningauthor{H.M. Wallach, S.T. Jensen, L. Dicker, and K.A. Heller}

\twocolumn[

\aistatstitle{An Alternative Prior Process for Nonparametric Bayesian
  Clustering}

\aistatsauthor{Hanna M. Wallach \And Shane T. Jensen}

\aistatsaddress{Department of Computer Science \\ University of Massachusetts Amherst \And Department of Statistics\\ The Wharton School, University of Pennsylvania } 

\aistatsauthor{Lee Dicker \And Katherine A. Heller}

\aistatsaddress{Department of Biostatistics \\ Harvard School of Public Health \And Engineering Department\\ University of Cambridge} ]

\begin{abstract}
Prior distributions play a crucial role in Bayesian approaches to
clustering. Two commonly-used prior distributions are the Dirichlet
and Pitman-Yor processes. In this paper, we investigate the predictive
probabilities that underlie these processes, and the implicit
``rich-get-richer'' characteristic of the resulting partitions. We
explore an alternative prior for nonparametric Bayesian
clustering---the uniform process---for applications where the
``rich-get-richer'' property is undesirable. We also explore the cost
of this process: partitions are no longer exchangeable with respect to
the ordering of variables. We present new asymptotic and
simulation-based results for the clustering characteristics of the
uniform process and compare these with known results for the Dirichlet
and Pitman-Yor processes. We compare performance on a real document
clustering task, demonstrating the practical advantage of the uniform
process despite its lack of exchangeability over orderings.
\end{abstract}

\section{Introduction}
\label{introduction}

Nonparametric Bayesian models provide a powerful and popular approach
to many difficult statistical problems, including document clustering
\cite[]{ZhaGhaYan05}, topic modeling~\cite[]{teh06hierarchical}, and
clustering motifs in DNA sequences~\cite[]{JenLiu08}. The key
assumption underlying nonparametric Bayesian models is the existence
of a set of random variables drawn from some unknown probability
distribution. This unknown probability distribution is itself drawn
from some prior distribution. The Dirichlet process is one such prior for
unknown probability distributions that has become ubiquitous in
Bayesian nonparametric modeling, as reviewed by \cite{MulQui04}. More
recently, \cite{PitYor97} introduced the Pitman-Yor process, a
two-parameter generalization of the Dirichlet process. These processes
can also be nested within a hierarchical
structure~\cite[]{TehJorBea06,Teh06}. A key property of any model
based on Dirichlet or Pitman-Yor processes is that the posterior
distribution provides a partition of the data into clusters, without
requiring that the number of clusters be pre-specified in advance.
However, previous work on nonparametric Bayesian clustering has paid
little attention to the implicit \emph{a priori} ``rich-get-richer''
property imposed by both the Dirichlet and Pitman-Yor process. As we
explore in section~\ref{priors}, this property is a fundamental
characteristic of partitions generated by these processes, and leads
to partitions consisting of a small number of large clusters, with
``rich-get-richer'' usage. Although ``rich-get-richer'' cluster usage
is appropriate for some clustering applications, there are others for
which it is undesirable. As pointed out by \cite{Wel06}, there exists
a need for alternative priors in clustering models.

In this paper, we explore one such alternative prior---the
\emph{uniform process}---which exhibits a very different set of
clustering characteristics to either the Dirichlet process or the
Pitman-Yor process. The uniform process was originally introduced by
\cite{QinMcCTho03} (page 438) as an \emph{ad hoc} prior for DNA motif
clustering.  However, it has received little attention in the
subsequent statistics and machine learning literature and its
clustering characteristics have remained largely unexplored. We
therefore compare the uniform process to the Dirichlet and Pitman-Yor
processes in terms of asymptotic characteristics
(section~\ref{asymptotics}) as well as characteristics for sample
sizes typical of those found in real clustering applications
(section~\ref{simulationstudy}). One fundamental difference between
the uniform process and the Dirichlet and Pitman-Yor processes is the
uniform process's lack of exchangeability over cluster
assignments---the probability $P(\bc)$ of a particular set of cluster
assignments $\bc$ is not invariant under permutations of those
assignments. Previous work on the uniform process has not acknowledged
this issue with respect to either inference or probability
calculations. We demonstrate that this lack of exchangeability is not
a significant problem for applications where a more balanced prior
assumption about cluster sizes is desired. We present a new Gibbs
sampling algorithm for the uniform process that is correct for a fixed
ordering of the cluster assignments, and show that while $P(\bc)$ is
not invariant to permuted orderings, it can be highly robust.

We also consider the uniform process in the context of a real text
processing application: unsupervised clustering of a set of documents
into natural, thematic groupings.  An extensive and diverse array of
models and procedures have been developed for this task, as reviewed
by \cite{AndFox07}. These approaches include nonparametric Bayesian
clustering using the Dirichlet process \cite[]{ZhaGhaYan05} and the
hierarchical Dirichlet process \cite[]{TehJorBea06}. Such
nonparametric models are popular for document clustering since the
number of clusters is rarely known \emph{a priori}, and these models
allow the number of clusters to be inferred along with the assignments
of documents to clusters. However, as we illustrate below, the
Dirichlet process still places prior assumptions on the clustering
structure: partitions will typically be dominated by a few very large
clusters, with overall ``rich-get-richer'' cluster usage. For many
applications, there is no \emph{a priori} reason to expect that this
kind of partition is preferable to other kinds of partitions, and in
these cases the uniform process can be a better representation of
prior beliefs than the Dirichlet process. We demonstrate that the
uniform process leads to superior document clustering performance
(quantified by the probability of unseen held-out documents under the
model) over the Dirichlet process using a collection of carbon
nanotechnology patents (section~\ref{application}).

\section{Predictive Probabilities for Clustering Priors}
\label{priors}

Clustering involves partitioning random variables $\bX = (X_1, \ldots,
X_N)$ into clusters. This procedure is often performed using a mixture
model, which assumes that each variable was generated by one of $K$ 
mixture components characterized by parameters $\Phi = \{ \phi_k \}_{k=1}^K$: 
\begin{eqnarray}
P(X_n \g\Phi)=\sum_{k=1}^K P(c_n \!=\! k)\,P(X_n \g \phi_k,
c_n \!=\! k),
\end{eqnarray}
where $c_n$ is an indicator variable such that $c_{n}=k$ if and only
if data point $X_n$ was generated by component $k$ with parameters
$\phi_k$. Clustering can then be characterized as identifying the set
of parameters responsible for generating each observation. The
observations associated with parameters $\phi_k$ are those $X_n$ for
which $c_n=k$. Together, these observations form cluster $k$. Bayesian
mixture models assume that the parameters $\Phi$ come from some
prior distribution $P(\Phi)$. Nonparametric Bayesian mixture models
further assume that the probability that $c_n=k$ is well-defined in
the limit as $K\rightarrow\infty$. This allows for more flexible
mixture modeling, while avoiding costly model comparisons in order to
determine the ``right'' number of clusters or components $K$.  From a
generative perspective, in nonparametric Bayesian mixture modeling,
each observation is assumed to have been generated by first selecting
a set of component parameters $\phi_k$ from the prior and then
generating the observation itself from the corresponding
component. Clusters are therefore constructed sequentially. The
component parameters responsible for generating a new observation are
selected using the \emph{predictive probabilities}---the conditional
distribution over component parameters implied by a particular choice
of priors over $\Phi$ and $c_n$. We next describe three priors---the
Dirichlet, Pitman-Yor, and uniform processes---using their predictive
probabilities. For notational convenience we define $\psi_n$ to be the
component parameters for the mixture component responsible for
observation $X_n$, such that $\psi_n = \phi_k$ when $c_n=k$.

\subsection{Dirichlet Process}

The Dirichlet process prior has two parameters: a \emph{concentration
  parameter} $\theta$, which controls the formation of new clusters,
and a \emph{base distribution} $G_0$. Under a Dirichlet process prior, the
conditional probability of the mixture component parameters
$\psi_{N+1}$ associated with a new observation $X_{N+1}$ given the
component parameters $\psi_1, \ldots, \psi_N$ associated with previous
observations $X_1, \ldots, X_N$ is a mixture of point masses at the
locations of $\psi_1, \ldots, \psi_N$ and the base distribution
$G_0$. Variables $X_n$ and $X_m$ are said to to belong to the same
cluster if and only if $\psi_n = \psi_m$.\footnote{Assuming a
  continuous $G_0$.} This predictive probability formulation
therefore sequentially constructs a partition, since observation
$X_{N+1}$ belongs to an existing cluster if $\psi_{N + 1} = \psi_n$
for some $n \leq N$ or a new cluster consisting only of $X_{N+1}$ if
$\psi_{N+1}$ is drawn directly from $G_0$. If $\phi_1, \ldots, \phi_K$
are the $K$ distinct values in $\psi_1, \ldots, \psi_N$ and $N_1,
\ldots, N_K$ are the corresponding cluster sizes (\emph{i.e.}, $N_k =
\sum_{n=1}^N {\rm I}\, (\psi_n = \phi_k)$, then
\begin{align}
&P(\psi_{N+1} \g \psi_1, \ldots,\psi_{N}, \theta, G_0) = \notag\\
&\qquad \begin{cases}
\frac{N_k}{N + \theta} & \psi_{N + 1} = \phi_k \in \{ \phi_1, \ldots, \phi_K \}\\
\frac{\theta}{N + \theta} & \psi_{N + 1} \sim G_0.
\end{cases}
\label{DPrule}
\end{align}
New observation $X_{N+1}$ joins existing cluster $k$ with probability
proportional to $N_k$ (the number of previous observations in that
cluster) and joins a new cluster, consisting of $X_{N+1}$ only, with
probability proportional to $\theta$. This predictive probability is
evident in the \emph{Chinese restaurant process} metaphor
\cite[]{aldous85exchangeability}.

The most obvious characteristic of the Dirichlet process predictive
probability (given by (2)) is the ``rich-get-richer'' property: the
probability of joining an existing cluster is proportional to the size
of that cluster. New observations are therefore more likely to join
already-large clusters. The ``rich-get-richer'' characteristic is also
evident in the \emph{stick-breaking} construction of the Dirichlet
process \cite[]{Set94,IshJam01}, where each unique point mass is
assigned a random weight. These weights are generated as a product of
Beta random variables, which can be visualized as breaks of a
unit-length stick. Earlier breaks of the stick will tend to lead to
larger weights, which again gives rise to the
``rich-get-richer'' property.

\subsection{Pitman-Yor Process}

The Pitman-Yor process \cite[]{PitYor97} has three parameters: a
concentration parameter $\theta$, a base distribution $G_0$, and a
\emph{discount parameter} $0 \le \alpha < 1$. Together, $\theta$ and
$\alpha$ control the formation of new clusters. The Pitman-Yor
predictive probability is
\begin{align}
&P(\psi_{N+1} \g \psi_1, \ldots, \psi_N, \theta, \alpha, G_0) = \notag \\
&\qquad \begin{cases}
\frac{N_k-\alpha}{N + \theta} & \psi_{N+1} = \phi_k \in \{ \phi_1,
\ldots, \phi_K \}\\
\frac{\theta + K \alpha}{N + \theta} & \psi_{N+1} \sim G_0.\label{PYrule}
\end{cases}
\end{align}
The Pitman-Yor process also exhibits the ``rich-get-richer''
property. However, the discount parameter $\alpha$ serves to reduce
the probability of adding a new observation to an existing cluster.
This prior is particularly well-suited to natural language processing
applications \cite[]{Teh06, Wallach08} because it yields power-law
behavior (cluster usage) when $0 < \alpha < 1$.

\subsection{Uniform Process}

Predictive probabilities (\ref{DPrule}) and (\ref{PYrule}) result in
partitions that are dominated by a few large clusters, since new
observations are more likely to be assigned to larger clusters. For
many tasks, however, a prior over partitions that induces more
uniformly-sized clusters is desirable. The uniform
process~\cite[]{QinMcCTho03,JenLiu08} is one such prior. The predictive
probability for the uniform process is given by
\begin{align}
&P(\psi_{N+1} \g \psi_1, \ldots, \psi_N, \theta, G_0) = \notag \\
&\qquad \begin{cases}
\frac{1}{K + \theta} & \psi_{N + 1} = \phi_k \in \{ \phi_1, \ldots,
\phi_K \}\\
\frac{\theta}{K + \theta} & \psi_{N+1} \sim G_0.\label{UPrule}
\end{cases}
\end{align}
The probability that new observation $X_{N+1}$ joins one of the
existing $K$ clusters is uniform over these clusters, and is unrelated to the cluster sizes. Although the uniform process has been
used previously for clustering DNA motifs
\cite[]{QinMcCTho03,JenLiu08}, its usage has otherwise been extremely
limited in the statistics and machine learning literature and its
theoretical properties have thus-far not been explored.

Constructing prior processes using predictive probabilities can imply
that the underlying prior results in nonexchangeability. If $\bc$
denotes a partition or set of cluster assignments for observations
$\bX$, then the partition is exchangeable if the calculation of the
full prior density of the partition $P(\bc)$ via the predictive
probabilities is unaffected by the ordering of the cluster
assignments. As discussed by \cite{Pit96} and \cite{Pit02}, most sequential processes will
fail to produce a partition that is exchangeable.  The Dirichlet
process and Pitman-Yor process predictive probabilities
((\ref{DPrule}) and (\ref{PYrule})) both lead to exchangeable
partitions. In fact,
their densities are special cases of ``exchangeable partition
probability functions'' given by \cite{IshJam03}.  \cite{GreRic01} and \cite{Wel06} discuss the relaxation of exchangeability in order to consider
alternative prior processes. The uniform process does not ensure
exchangeability: the prior probability $P(\bc)$ of a particular set of
cluster assignments $\bc$ is not invariant under permutation of those
cluster assignments. However, in section~\ref{exchangeability}, we
demonstrate that the nonexchangeability implied by the uniform process
is not a significant problem for real data by showing that $P(\bc)$ is
robust to permutations of the observations and hence cluster
assignments.

\section{Asymptotic Behavior}
\label{asymptotics}

In this section, we compare the three priors implied by predictive
probabilities (\ref{DPrule}), (\ref{PYrule}) and (\ref{UPrule}) in
terms of the asymptotic behavior of two partition characteristics: the
number of clusters $K_N$ and the distribution of cluster sizes $\bH_N
= (H_{1,N}, H_{2,N}, \ldots, H_{N,N})$ where $H_{M,N}$ is the number
of clusters of size $M$ in a partition of $N$ observations.  We begin
by reviewing previous results for the Dirichlet and Pitman-Yor
processes, and then present new results for the uniform process.

\subsection{Dirichlet
  Process} \label{DP_asymptotic}

As the number of observations $N \to \infty$, the expected number of
unique clusters $K_N$ in a partition is
\begin{equation}
\mathbb{E}\,(K_N\g{\rm DP}) = \sum_{n = 1}^N \frac{\theta}{n -
  1 + \theta}  \,\simeq\, \theta \log N. \label{DPmom2}
\end{equation}
The expected number of clusters of size $M$ is
\begin{equation}
\lim_{N \to \infty} \mathbb{E}\,(H_{M,N}\g{\rm DP}) = 
\frac{\theta}{M}.\label{DPmom1}
\end{equation}
This well-known result \cite[]{ArrBarTav03} implies that as $N \to
\infty$, the expected number of clusters of size $M$ is inversely
proportional to $M$ regardless of the value of $\theta$. In other words, in
expectation, there will be a small number of large clusters and \emph{vice versa}.

\subsection{Pitman-Yor Process} \label{PY_asymptotic}

\cite{Pit02} showed that as $N \to \infty$, the expected number
of unique clusters $K_N$ in a partition is
\begin{equation}
\mathbb{E}\,(K_N \g{\rm PY}) \approx \frac{\Gamma(1 + \theta)}{
  \alpha\Gamma(\alpha +
\theta)} \,N^{\alpha}. \label{PYmom2}
\end{equation}
Pitman's result can also be used to derive the expected number of
clusters of size $M$ in a partition:
\begin{equation}
\mathbb{E}\,(H_{M,N} \g{\rm PY}) \approx \frac{\Gamma(1 +
  \theta) \prod_{m=1}^{M-1} (m - \alpha)}{\Gamma(\alpha +
\theta)\,  M!} N^{\alpha}.  \label{PYmom4}
\end{equation}

\subsection{Uniform Process}\label{UN_asymptotic}

Previous literature on the uniform process does not contain any
asymptotic results. We therefore present the following novel result
for the expected number of unique clusters $K_N$ in a partition as $N
\to \infty$:
\begin{equation}
\mathbb{E}\,(K_N\g{\rm UP}) \approx \sqrt{2 \theta} \cdot
N^{\frac{1}{2}}. \label{UNmom2}
\end{equation}
A complete proof is given in the supplementary materials. In
section~\ref{simulationstudy}, we also present simulation-based
results that suggest the following conjecture for the expected number
of clusters of size $M$ in a partition:
\begin{equation}
\mathbb{E}\,(H_{M,N} \g {\rm UP}) \approx \theta.  \label{UNmom4}
\end{equation}
This result corresponds well to the intuition underlying the uniform
process: observations are \emph{a priori} equally likely to join any
existing cluster, regardless of size.

\subsection{Summary of Asymptotic Results}

The distribution of cluster sizes for the uniform process is
dramatically different to that of either the Pitman-Yor or Dirichlet
process, as evidenced by the results above, as well as the
simulation-based results in section~\ref{simulationstudy}. The uniform
process exhibits a uniform distribution of cluster sizes. Although the
Pitman-Yor process can be made to behave similarly to the uniform
process in terms of the expected number of clusters (by varying
$\alpha$, as described below), it cannot be configured to exhibit a
uniform distribution over cluster sizes, which is a unique aspect of
the uniform process.

Under the Dirichlet process, the expected number of clusters grows
logarithmically with the number of observations $N$. In contrast,
under the uniform process, the expected number of clusters grows with
the square root of the number of observations $N$. The Pitman-Yor
process implies that the expected number of clusters grows at a rate
of $N^\alpha$. In other words, the Pitman-Yor process can lead to a
slower or faster growth rate than the uniform process, depending on
the value of the discount parameter $\alpha$. For $\alpha = 0.5$, the
expected number of clusters grows at the same rate for both the
Pitman-Yor process and the uniform process.

\section{Simulation Comparisons: Finite $N$}
\label{simulationstudy}

The asymptotic results presented in the previous section are not
necessarily applicable to real data where the finite number of
observations $N$ constrains the distribution of cluster sizes,
$\sum_{M} M \cdot H_{M,N} = N$.  In this section, we appraise the
finite sample consequences for the Dirichlet, Pitman-Yor, and uniform
processes via a simulation study. For each of the three processes, we
simulated 1000 independent partitions for various values of sample
size $N$ and concentration parameter $\theta$, and calculated the
number of clusters $K_N$ and distribution of cluster sizes $\bH_N$ for
each of the partitions.

\subsection{Number of Clusters $K_N$}

\begin{figure}[t]
\centering
\includegraphics[width=3.35in,height=2in]{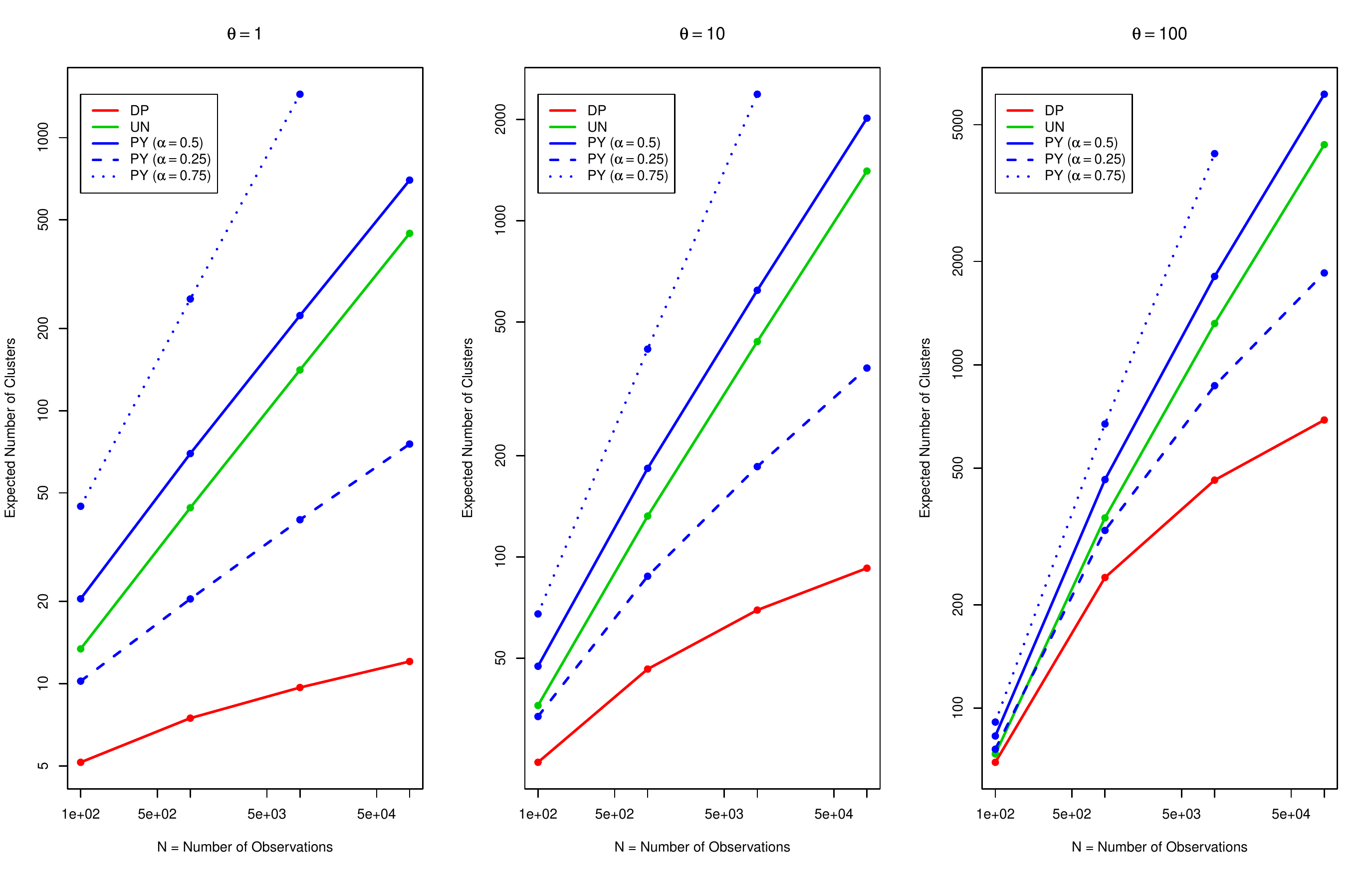}
\caption{Expected number of clusters $\hat{K}_N$ versus sample size
  $N$ for different $\theta$. Axes are on a log
  scale.}\label{numberofclusters}
\vspace{-0.4cm}
\end{figure}

\begin{figure*}[t]
\centering
\includegraphics[width=4in,height=2.25in]{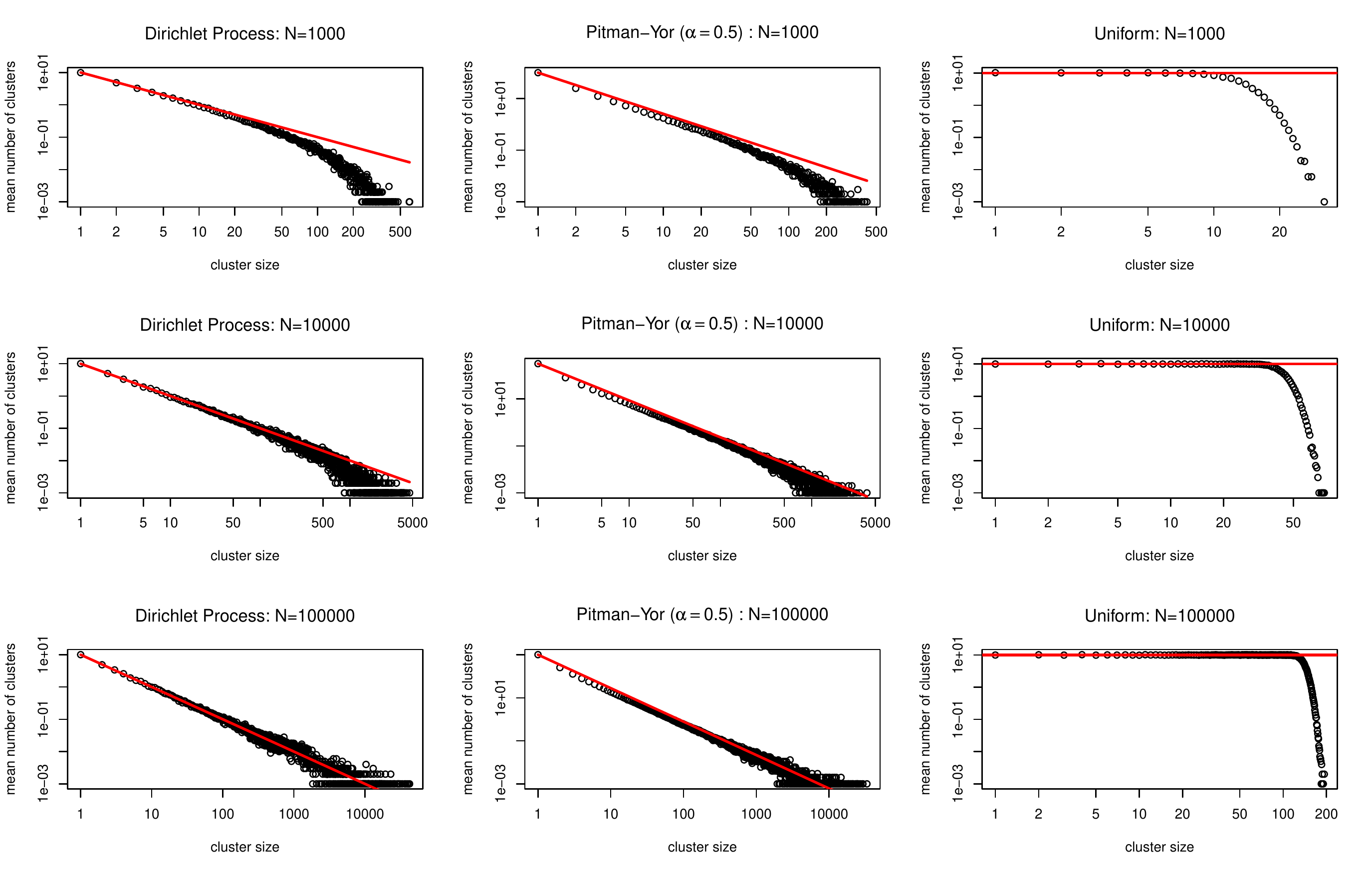}
\caption{Cluster sizes $H_{M,N}$ as a function of $M$ for different
  values of $N$ for the Dirichlet, Pitman-Yor, and uniform processes. Data are plotted on
  a log-log scale and the red lines indicate the asymptotic
  relationships.  Each point is the average number of clusters (across 1000 simulated partitions) of a particular cluster size.}\label{simclustersizes}
\vspace{-0.4cm}
\end{figure*}

In figure~\ref{numberofclusters}, we examine the relationship between
the number of observations $N$ and the average number of clusters
$\hat{K}_N$ (averaged over the 1000 simulated partitions). For $\alpha
= 0.5$, the Pitman-Yor process exhibits the same rate of growth of
$\hat{K}_N$ as the uniform process, confirming the equality suggested
by (\ref{PYmom2}) and (\ref{UNmom2}) when $\alpha = 0.5$.  As
postulated in section~\ref{PY_asymptotic}, the Pitman-Yor process can
exhibit either slower (\emph{e.g.}, $\alpha=0.25$) or faster (\emph{e.g.},
$\alpha=0.75$) rates of growth of $\hat{K}_N$ than the uniform
process. The rate of growth of $\hat{K}_N$ for the Dirichlet process
is the slowest, as suggested by (\ref{DPmom2}).

\subsection{Distribution of Cluster Sizes}

In this section, we examine the expected distribution of cluster sizes
under each process. For brevity, we focus only on concentration
parameter $\theta = 10$, though the same trends are observed for other
values of $\theta$. Figure~\ref{simclustersizes} is a plot of
$\hat{H}_{M,N}$ (the average number of clusters of size $M$) as a
function of $M$. For each process, $\hat{H}_{M,N}$ was calculated as
the average over the 1000 simulated independent partitions of
$H_{M,N}$ under that process. The red lines indicate the asymptotic
relationships, \emph{i.e.}, (\ref{DPmom1}) for the Dirichlet process,
(\ref{PYmom4}) for the Pitman-Yor process, and (\ref{UNmom4}) for the
uniform process.

The results in figure~\ref{simclustersizes} demonstrate that the
simulated distribution of cluster sizes for the uniform process is
quite different to the simulated distributions of clusters sizes for
either the Dirichlet or Pitman-Yor processes. It is also interesting
to observe the divergence from the asymptotic relationships due to the
finite sample sizes, especially in the case of small $N$
(\emph{e.g.}, $N = 1000$).

\section{Exchangeability}
\label{exchangeability}

As mentioned in section~\ref{priors}, the uniform process does not
lead to exchangeable partitions. Although the exchangeability of the
Dirichlet and Pitman-Yor processes is desirable, these clustering
models also exhibit the ``rich-get-richer'' property. Applied
researchers are routinely forced to make assumptions when modeling
real data. Even though the use of exchangeable priors can provide many
practical advantages for clustering tasks, exchangeability itself is
one particular modeling assumption, and there are situations in which
the ``rich-get-richer'' property is disadvantageous. In reality, many
data generating processes are not exchangeable, \emph{e.g.}, news
stories are published at different times and therefore have an
associated temporal ordering. If one is willing to make an
exchangeability assumption, then the Dirichlet process prior is a
natural choice. However, it comes with additional assumptions about
the size distribution of clusters. These assumptions will be
reasonable in certain situations, but less reasonable in others. It
should not be necessary to restrict applied researchers to
exchangeable models, which can impose other undesired assumptions,
when alternatives do exist. The uniform process sacrifices the
exchangeability assumption in order to make a more balanced prior
assumption about cluster sizes.

In this section, we explore the lack of exchangeability of the uniform
process by first examining, for real data, the extent to which
$P(\bc)$ is affected by permuting the observations. For any particular
ordering of observations $\bX = (X_1, \ldots, X_N)$, the joint
probability of the corresponding cluster assignments $\bc$ is
\begin{equation}
P(\bc \g \textrm{ordering $1, \ldots, N$}) = \prod_{n=1}^N P(c_n \g
\bc_{<n})
\end{equation}
where ``$\bc_{<n}$'' denotes the cluster assignments for observations
$X_1, \ldots, X_{n-1}$ and $P(c_n \g \bc_{<n})$ is given by
(\ref{UPrule}). Clearly, exhaustive evaluation of $P(\bc)$ for all
possible orderings (permutations of observations) is not possible for
realistically-sized data sets. However, we can evaluate the robustness
of $P(\bc)$ to different orderings as follows: for any given partition
$\bc$ (set of cluster assignments), we can compute the standard
deviation of $\log P(\bc)$ over multiple different orderings of the
observations. This  ``between-ordering SD''
gives an estimate of the degree to which the ordering of observations
affects $P(\bc)$ for a particular partition. For any given ordering of
observations, we can also compute the standard deviation of $\log{P(\bc)}$ over multiple
different partitions (realizations of $\bc$) inferred using the
Gibbs sampling algorithm described below. This ``between-partition SD''  gives an
estimate of the variability of inferred partitions for a fixed
ordering. 

\begin{figure}[t]
\vspace{-0.5cm}
\centering
\rotatebox{270}{\includegraphics[width=3.5in,height=2.25in,angle=90]{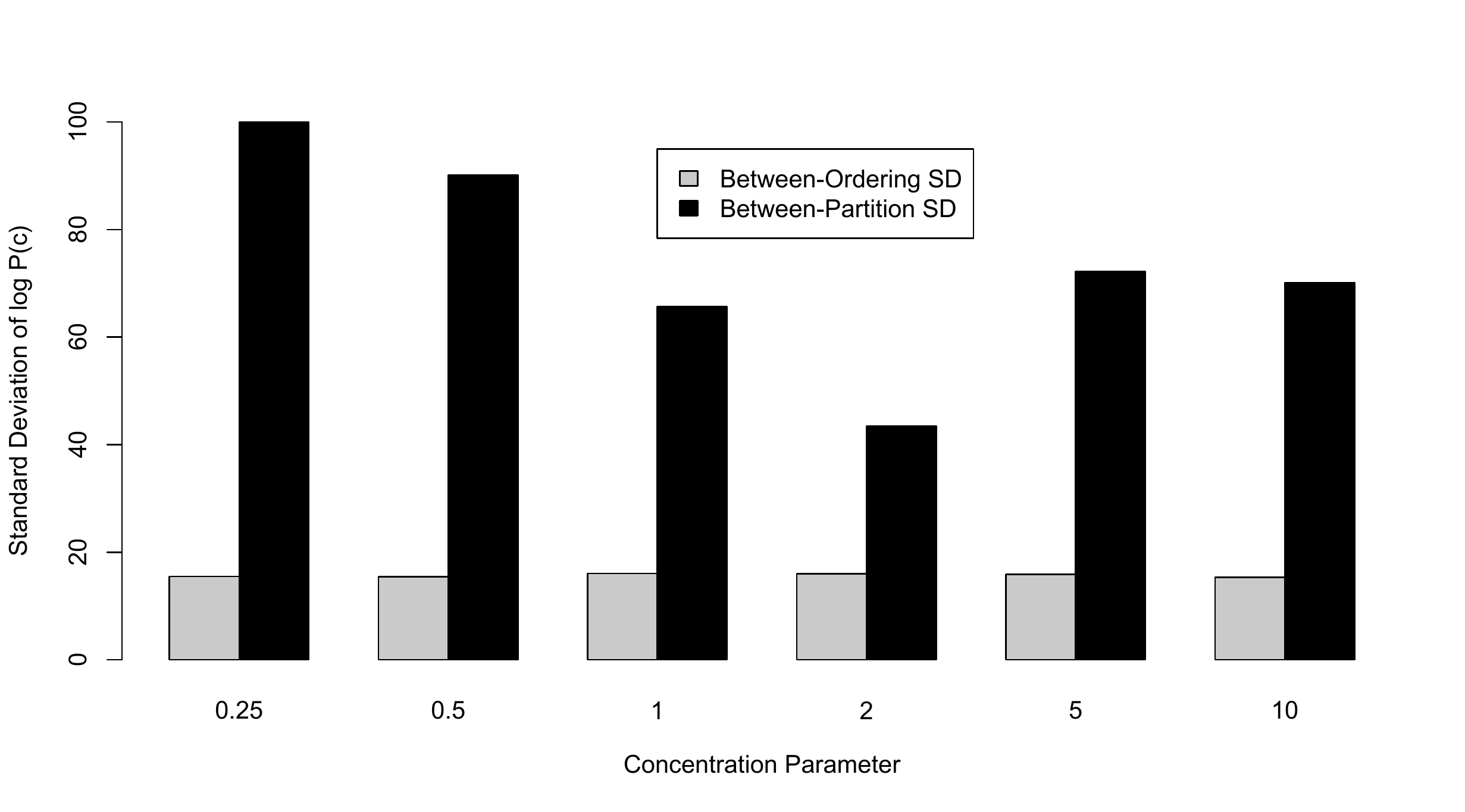}}
\caption{Comparison of the ``Between-Partition SD'' and the ``Between-Ordering
  SD'' (averaged over different inferred partitions) for the uniform
  process with six different values of concentration parameter
  $\theta$.}\label{orderingfig}
\vspace{-0.4cm}
\end{figure}

Figure~\ref{orderingfig} shows the ``between-ordering SD'' and the
``between-partition SD'' for partitions of 1000 carbon nanotechnology
patent abstracts (see next section), obtained using five Gibbs
sampling chains and 5000 orderings of the data with different values
of $\theta$.  The variability between orderings is considerably
smaller than the variability between partitions, suggesting that
uniform process clustering results are not significantly sensitive to
different orderings. These results are encouraging for applications
where one is willing to sacrifice exchangeability over orderings in
favor of a more balanced prior assumption about cluster sizes.

\section{Document Clustering Application}\label{application}

In this section, we compare the Dirichlet process and the uniform
processes on the task of clustering real
documents---specifically, the
abstracts of 1200 carbon nanotechnology patents.  Dirichlet processes
have been used as the basis of many approaches to document clustering 
including those of \cite{ZhaGhaYan05}, \cite{ZhuGhaLaf05} and
\cite{Wal08}. In practice, however, there is often little
justification for the \emph{a priori} ``rich-get-richer'' property
exhibited by the Dirichlet process.

We consider a nonparametric word-based mixture model where documents
are clustered into groups on the basis of word occurrences. The model
assumes the following generative process: The tokens $\bw_d$ that
comprise each document, indexed by $d$, are drawn from a
document-specific distribution over words $\bphi_d$, which is itself
drawn from a document-specific Dirichlet distribution with base
distribution $\bn_d$ and concentration parameter $\beta$. The
document-specific base distribution is obtained by selecting a cluster
assignment from the uniform process.  If an existing cluster is
selected, then $\bn_d$ is set to the cluster-specific distribution
over words for that cluster. If a new cluster is selected, then a new
cluster-specific distribution over words is drawn from $G_0$, and
$\bn_d$ is set to that distribution:
\begin{eqnarray}
c_d \g c_{< d} &\sim&
\begin{cases} \frac{1}{d-1 + \theta} & c_d = k \in {1, \ldots,
    K}\\
\frac{\theta}{d-1 + \theta} & c_d = k_{\textrm{new}}
\end{cases}\\
% \textrm{derived equation (\ref{UPrule})}\\
\bn_k &\sim& G_0\\
\bphi_d &\sim& \textrm{Dir}\,(\bphi_d \g \bn_{c_d}, \beta)\\
\bw_d &\sim& \textrm{Mult}\,(\bphi_d),
\end{eqnarray}
where $c_d$ is the cluster assignment for the $d^{\textrm{th}}$
document. Finally, $G_0$ is chosen to be a hierarchical Dirichlet
distribution: $G_0 = \textrm{Dir}\,(\bn_{c} \g \beta_1 \bn)$, where
$\bn \sim \textrm{Dir}\,(\bn \g \beta_0 \bu)$. This model captures the
fact that documents in different clusters are likely to use different
vocabularies, yet allows the word distribution for each document
to vary slightly from the word distribution for the cluster to which
that document belongs.

The key consequence of using either a Dirichlet or uniform process
prior is that the latent variables $\bn_d$ are partitioned into $C$
clusters where the value of $C$ does not need to be pre-specified and fixed.  The
vector $\bc$ denotes the cluster assignments for the documents: $c_d$
is the cluster assignment for document $d$.  Given a set of observed
documents
$\mathcal{W} = \{\bw_d \}_{d=1}^D$, Gibbs
sampling \cite[]{geman84stochastic} can be used to infer the latent cluster
assignments $\bc$. Specifically, the cluster assignment $c_d$ for
document $d$ can be resampled from
\begin{align}
&P(c_d \g \bc_{\setminus d}, \bw, \theta) \propto \notag\\
&\qquad P(c_d \g
\bc_{\setminus d}, \theta) \, \cdot \, P(\bw_d \g c_d,
\bc_{\setminus d}, \mathcal{W}_{\setminus d},
\bbeta),\label{gibbssampling}
\end{align}
where $\bc_{\setminus d}$ and $\mathcal{W}_{\setminus d}$ denote the
sets of clusters and documents, respectively, excluding document $d$.
The vector $\bbeta = (\beta, \beta_1, \beta_0)$ represents the concentration
parameters in the model, which can be inferred from $\mathcal{W}$ using slice
sampling \cite[]{neal03slice}, as described by \cite{Wal08}. The
likelihood component of (\ref{gibbssampling}) is
\begin{align}
& P(\bw_d \g c_d,
\bc_{\setminus d}, \mathcal{W}_{\setminus d}, \bbeta) = \notag \\
&\qquad \prod_{n=1}^{N_d} \frac{N_{w_n|d}^{<d,n} + \beta\,
  \frac{N_{w_n|c_d}^{<d,n} + \beta_1\, \frac{N_{w_n}^{<d,n} +
      \beta_0\,
      \frac{1}{W}}{\sum_w N_w^{<d,n} + \beta_0}}{\sum_w
    N_{w|c_d}^{<d,n} +
    \beta_1}}{\sum_w N_{w|d}^{<d,n} + \beta},
\end{align}
where the superscript ``$< d,n$'' denotes a quantity including data
from documents $1, \ldots, d$ and positions $1, \ldots, n-1$ only for
document $d$. $N_{w|d}$ is the number of times word type $w$ occurs in
document $d$, $N_{w|c_d}$ is the number of times $w$ occurs in cluster
$c_d$, and $N_w$ is the number of times $w$ occurs in the entire
corpus.

The conditional prior probability $P(c_d \g \bc_{\setminus d},
\theta)$ can be constructed using any of the predictive probabilities
in section~\ref{priors}. For brevity, we focus on the 
(commonly-used) Dirichlet process and the uniform process. For the
Dirichlet process, the conditional prior probability is given by
(\ref{DPrule}). Since the uniform process lacks exchangeability over
observations, we condition upon an arbitrary ordering of the
documents, \emph{e.g.}, $1, \dots, D$. The conditional prior of $c_d$ given
$\bc_{\setminus d}$ is therefore
\begin{align}
&P(c_d \g \bc_{\setminus d}, \theta, \textrm{ ordering $1, \ldots, D$})
\propto \notag\\
&\qquad P(c_d \g c_1,\ldots,c_{d-1}, \theta) \notag \\
&\qquad \prod_{m=d+1}^{D} P(c_m
\g c_1,\ldots,c_{m-1}, \theta),
\end{align}
where $P(c_d \g c_1,\ldots,c_{d-1}, \theta)$ is given by
(\ref{UPrule}).  The latter terms propagate the value of $c_d$ to the
cluster assignments $c_{d+1}, \ldots, c_D$ for the documents that
follow document $d$ in the chosen ordering. With this definition of
the conditional prior, the Gibbs sampling algorithm is a correct
clustering procedure for $\mathcal{W}$, conditioned on the arbitrarily
imposed ordering of the documents.

We compare the Dirichlet and uniform process priors by using the model
(with each prior) to cluster 1200 carbon nanotechnology patent
abstracts. For each prior, we use Gibbs sampling and slice sampling to
infer cluster assignments $\bc^{\textrm{train}}$ and $\bbeta$ for a
subset $\mathcal{W}^{\textrm{train}}$ of 1000 ``training''
abstracts. Since the results in section~\ref{exchangeability} indicate
that the variability between partitions is greater than the
variability between orderings, we use a single ordering of
$\mathcal{W}^{\textrm{train}}$ and perform five runs of the Gibbs
sampler. To provide insight into the role of $\theta$, we compare
results for several $\theta$ values. We evaluate predictive
performance by computing the probability of a held-out set
$\mathcal{W}^{\textrm{test}}$ of 200 abstracts given each run from
the trained model. We compute $\log{P(\mathcal{W}^{\textrm{test}} \g
  \mathcal{D}^{\textrm{train}}, \theta, \bbeta)} =
\log{\sum_{\bc^{\textrm{test}}} P(\mathcal{W}^{\textrm{test}},
  \bc^{\textrm{test}} \g \mathcal{D}^{\textrm{train}}, \theta,
  \bbeta)}$, where $\mathcal{D}^{\textrm{train}} =
(\mathcal{W}^{\textrm{train}}, \bc^{\textrm{train}})$ and the sum over
$\bc^{\textrm{test}}$ is approximated using a novel variant of
\cite[]{wallach09evaluation}'s ``left-to-right'' algorithm (see
supplementary materials). We average this quantity over runs of the
Gibbs sampler for $\mathcal{W}^{\textrm{train}}$, runs of the
``left-to-right'' algorithm, and twenty permutations of the held-out
data $\mathcal{W}^{\textrm{test}}$.

\begin{figure*}[t]
\vspace{-0.25cm}
\centering
\rotatebox{270}{\includegraphics[width=4.5in,height=2.25in,angle=90]{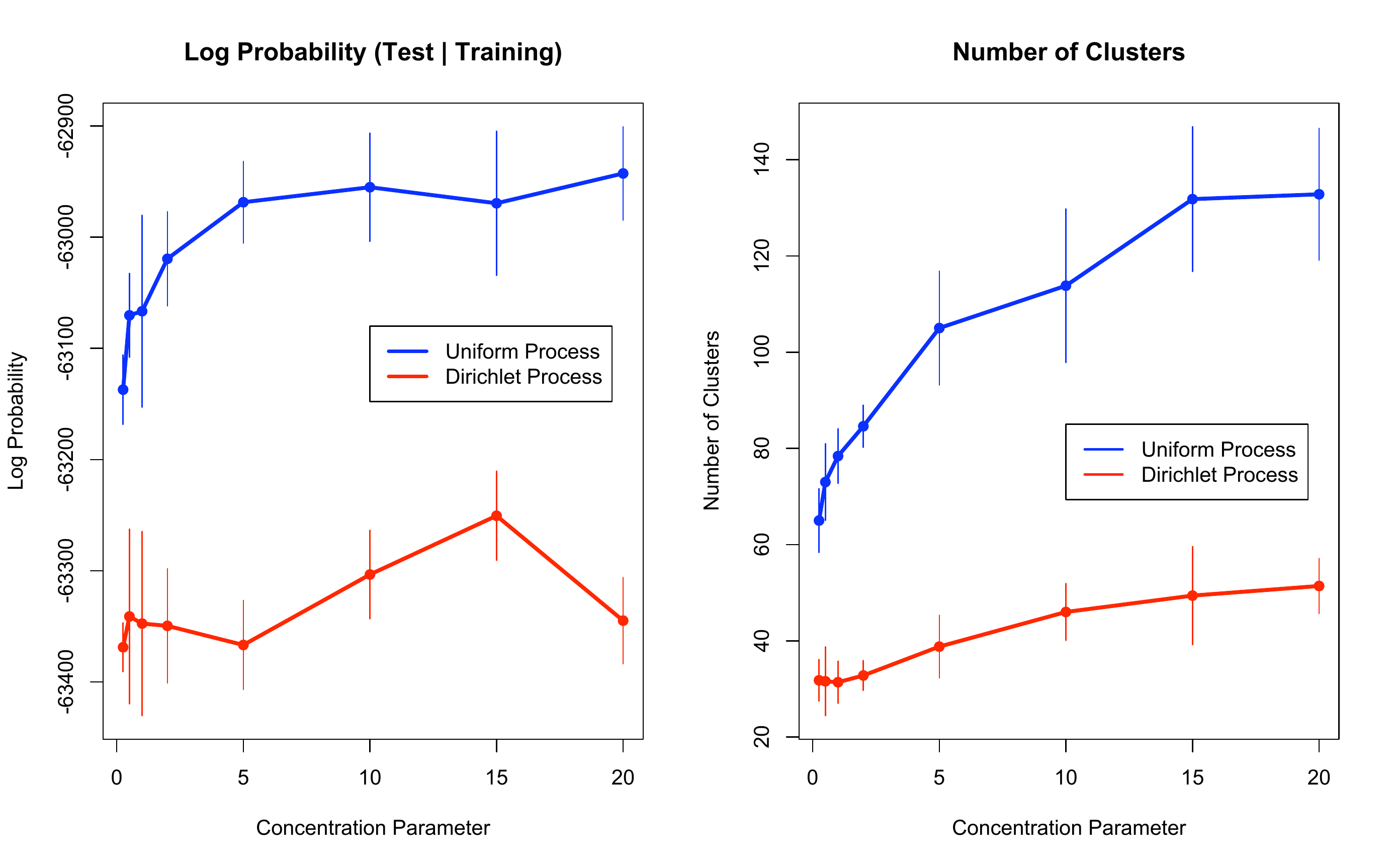}}
\caption{Left: Average
  log
  probability of held-out data given the trained model. Vertical lines indicate one SD across runs of the
  Gibbs sampler for $\mathcal{W}^{\textrm{train}}$, runs of the
  evaluation algorithm and (for the uniform process) twenty
  permutations of the held-out data. Right: The average number of
  clusters
  in a representative partition from each trained model. Vertical
  lines indicate one SD across runs of the Gibbs
  sampler for $\mathcal{W}^{\textrm{train}}$.}\label{resultsfig}
\vspace{-0.4cm}
\end{figure*}

The left-hand plot of figure~\ref{resultsfig} compares the Dirichlet
and uniform processes in terms of $\log{P(\mathcal{W}^{\textrm{test}}
  \g \mathcal{D}^{\textrm{train}}, \theta, \bbeta)}$. Regardless of
the value of concentration parameter $\theta$, the model based on the uniform process leads to
systematically higher held-out probabilities than the model based on
the Dirichlet process. In other words, the uniform process provides a
substantially better fit for the data in this application.  The
right-hand plot of figure~\ref{resultsfig} compares the Dirichlet and
uniform processes in terms of the average number of clusters in a
representative partition obtained using the Gibbs sampler. The uniform
process leads to a greater number of clusters than the Dirichlet
process for each value of $\theta$.  This is not surprising given the
theoretical results for the {\it a priori} expected cluster sizes
(section~\ref{asymptotics}) and the fact that the choice of clustering
prior is clearly influential on the posterior distribution in this
application.

\section{Discussion}\label{discussion}

The Dirichlet and Pitman-Yor processes both exhibit a
``rich-get-richer'' property that leads to partitions with a small
number of relatively large clusters and \emph{vice versa}. This
property is seldom fully acknowledged by practitioners when using
either process as part of a nonparametric Bayesian clustering
model. We examine the uniform process prior, which does not exhibit
this ``rich-get-richer'' property. The uniform process prior has
received relatively little attention in the statistics literature to
date, and its clustering characteristics have remained largely
unexplored. We provide a comprehensive comparison of the uniform
process with the Dirichlet and Pitman-Yor processes, and present a new
asymptotic result for the square-root growth of the expected number of
clusters under the uniform process. We also conduct a simulation study
for finite sample sizes that demonstrates a substantial difference in
cluster size distributions between the uniform process and the
Pitman-Yor and Dirichlet processes. Previous work on the uniform
process has ignored its lack of exchangeability. We present new
results demonstrating that although the uniform process is not
invariant to permutations of cluster assignments, it is highly
robust. Finally, we compare the uniform and Dirichlet processes on a
real document clustering task, demonstrating superior predictive
performance of the uniform process over the Dirichlet process.

\subsubsection*{Acknowledgements}

{\small
This work was supported in part by the Center for Intelligent
Information Retrieval, in part by CIA, NSA and NSF under NSF grant
\#IIS-0326249, and in part by subcontract \#B582467 from Lawrence
Livermore National Security, LLC, prime contractor to DOE/NNSA
contract \#DE-AC52-07NA27344. Any opinions, findings and conclusions or
recommendations expressed in this material are the authors' and do not
necessarily reflect those of the sponsor.
}

%Use unnumbered third level headings for the acknowledgements.  All acknowledgements go at the end of the paper.

%\subsubsection*{References}

\end{document}